\documentclass[aps, twocolumn, pra, amsmath, superscriptaddress,preprintnumbers]{revtex4-1}
\usepackage{graphicx}
\usepackage{dcolumn}
\usepackage{bm}
\usepackage{amsmath}
\usepackage{subfigure}
\usepackage{amsfonts}
\usepackage{xcolor}
\usepackage{appendix}
\usepackage{multirow}
\usepackage{booktabs}
\usepackage{tabularx}
\usepackage[colorlinks,
            linkcolor=blue,
            anchorcolor=red,
            citecolor=blue
            ]{hyperref}
\usepackage{amsthm}
\usepackage{setspace}
\usepackage{lineno}

\begin{document}
%\linenumbers

\title{Algebraic structure of path-independent quantum control}
\author{Wen-Long Ma}
\email{wenlongma@semi.ac.cn}
\affiliation{State Key Laboratory for Superlattices and Microstructures, Institute of Semiconductors, Chinese Academy of Sciences, Beijing 100083, China}
\affiliation{Pritzker School of Molecular Engineering, University of Chicago, Illinois 60637, USA}
\author{Shu-Shen Li}
\affiliation{State Key Laboratory for Superlattices and Microstructures, Institute of Semiconductors, Chinese Academy of Sciences, Beijing 100083, China}
\author{Liang Jiang}
%\email{liang.jiang@uchicago.edu}
\affiliation{Pritzker School of Molecular Engineering, University of Chicago, Illinois 60637, USA}

\date{\today }

\begin{abstract}
Path-independent (PI) quantum control has recently been proposed to integrate quantum error correction and quantum control [Phys. Rev. Lett. \textbf{125}, 110503 (2020)], achieving fault-tolerant quantum gates against ancilla errors. Here we reveal the underlying algebraic structure of PI quantum control. The PI Hamiltonians and propagators turn out to lie in an algebra isomorphic to the ordinary matrix algebra, which we call the PI matrix algebra. The PI matrix algebra, defined on the Hilbert space of a composite system (including an ancilla system and a central system), is isomorphic to the matrix algebra defined on the Hilbert space of the ancilla system. By extending the PI matrix algebra to the Hilbert-Schmidt space of the composite system, we provide an exact and unifying condition for PI quantum control against ancilla noise.

\end{abstract}

\maketitle

%%%%%%%%%%%%%%%%%%%%%%%%%%%%%%%%%%%%%%%%%%%%%%%%%%%%%%%%%%%%%%%%%%%%%%%%%%%%%%%%
%%%%%%%%%%%%%%%%%%%         INTRODUCTION  %%%%%%%%%%%%%%%%%%%%%%%%%%%%%%%%%%%%%%%%%%%%%%%%%
%%%%%%%%%%%%%%%%%%%%%%%%%%%%%%%%%%%%%%%%%%%%%%%%%%%%%%%%%%%%%%%%%%%%%%%%%%%%%%%%
To build a powerful quantum computer, the constituting quantum devices should have both good coherence and reliable universal control \cite{Nielsen2010,Glaser2015,Vandersypen2005}, which are often contradicting requirements. To have good coherence, we can choose the physical systems (called central system) well isolated from their noisy environment, such as the superconducting cavities \cite{Devoret2013,Schuster2007,Leghtas2013} and nuclear spins \cite{Kane1998,Dutt2007,Maurer2012,Liu2017}. The central system coherence can be further improved by either passive or active protection, such as dynamical decoupling  \cite{Viola1999,Liu2013,Taminiau2014}, decoherence-free subspace (subsystem) \cite{Lidar1998,Bacon1999} and quantum error correction \cite{Shor1995,Knill1997,Ofek2016,Waldherr2014}. However, as we try to realize a central system with nearly perfect coherence, it also becomes more difficult to process the quantum information in the central system, since reliable and fast control needs strong coupling with the outside world. One possible solution is to introduce an ancilla system, such as transmon qubits \cite{Koch2007,Krastanov2015,Heeres2015} and electron spins \cite{Dutt2007,Maurer2012}, which are relatively easily to control. However, since the ancilla system typically suffers more decoherence than the central system, the fidelity of the ancilla-assisted quantum operations is seriously limited by the ancill noise.  Therefore, it is crucial to develop quantum control protocols that are fault-tolerant against ancilla errors, therefore boosting the performance of ancilla-assisted quantum operations by largely suppressing ancilla errors.

Recently, we have proposed a general class of fault-tolerant quantum gates against ancilla errors, called path-independent (PI) quantum gates \cite{Ma2020}. The PI gates integrate quantum control and quantum error correction, guiding the design of hardware-efficient robust quantum operations against ancilla errors. The main feature of PI gates is that for given initial and final ancilla states, the central system undergoes a unitary gate independent of the specific ancilla path induced by control drives and ancilla error events. For specific final ancilla states, the desired unitary gate is implemented; for all other final ancilla states, the gate fails but the encoded information can be restored. Hence, we can perform conditional operation until the gate succeeds. A special class of PI gates are the previously proposed error-transparent gates for quantum error correction (QEC) codes \cite{Vy2013,Kapit2018,Ma2020b}, with the error syndromes corresponding to the ancilla states. Another important example of PI gates is the photon-number selective arbitrary phase (SNAP) gates in superconducting circuits \cite{Krastanov2015,Heeres2015}, which has recently been experimentally demonstrated with the gate fidelity significantly improved by the PI design \cite{Reinhold2020}. However, the general formalism in \cite{Ma2020} depends on a Dyson expansion of the Liouville superoperator, while the underling mathematical structure of path-independence criteria remains elusive.

In this paper, we provide deep insights on PI gates by uncovering the underlying algebraic structure, which we call the \textit{PI matrix algebra}. The PI matrix algebra is defined on a composite system containing the ancilla and central systems, but \emph{isomorphic} to the ordinary matrix algebra defined on the ancilla system alone. The PI control Hamiltonians and propagators found in \cite{Ma2020} belong to the PI matrix algebra. The path independence against ancilla error paths is connected to the multiplication operation of the PI matrix algebra. We also extend the PI matrix algebra to the Hilbert-Schmidt (HS) space of the composite system, and find a general class of PI superoperators. The path-independence criteria in \cite{Ma2020} can therefore be reformulated in an exact and unifying way.

\textit{Definition of PI matrix algebra}.---An algebra is a vector space together with a multiplication operation. A typical example is the ordinary matrix algebra $\mathcal{M}_{\rm A}$ for a $d_A$-level quantum system (ancilla system) with an orthonormal basis $\{|m\rangle_{A}\}_{m=1}^{d_A}$. $\mathcal{M}_{\rm A}$ is a vector space over the complex field $\mathbb{C}$ with the basis $\mathcal{B}_{A}=\{|m\rangle_A\langle n|\}_{m,n=1}^{d_A}$ and the multiplication operation
\begin{align}\label{def-product2}
    |a\rangle_A\langle b|c\rangle_A\langle d|=\delta_{bc}|a\rangle_A\langle d|,
\end{align}
with $a,b,c,d\in[1,d_A]$. Any quantum operator for the ancilla can be represented by a vector in $\mathcal{M}_{\rm A}$, and the product of any two operators can be obtained from the the above multiplication for base vectors.

Now we introduce another $d_B$-dimensional system (central system) with an orthonormal basis $\{|j\rangle_B\}_{j=1}^{d_B}$. For the composite system containing the ancilla and the central systems, we can define a matrix algebra isomorphic to $\mathcal{M}_{\rm A}$. The formal definition is as follows.

\textbf{Definition 1}: \textbf{PI matrix algebra}. Consider a vector space with the basis $\mathcal{B}_{AB}=\{|m\rangle_A\langle n|\otimes U_{mn}\}_{m,n=1}^{d_A}$ over the complex field $\mathbb{C}$. Here $\{U_{mn}\}$ is a discrete set of unitary operators on the central system satisfying $U_{me}U_{en}=U_{mn}$ with $m,e,n\in[1,d_A]$, from which we can derive $U_{mm}=\mathbb{I}_B$ being the identity operation on the central system and $U_{mn}=U^{\dagger}_{nm}$ with ${\dagger}$ denoting the Hermitian conjugation. The multiplication operation of the base vectors in $\mathcal{B}_{AB}$ is
\begin{align}\label{def-product}
    (|a\rangle_A\langle b|\otimes U_{ab})\cdot(|c\rangle_A\langle d|\otimes U_{cd})=\delta_{bc}|a\rangle_A\langle d|\otimes U_{ad},
\end{align}
with $a,b,c,d\in[1,d_A]$. Define this vector space with the multiplication operation in Eq. (\ref{def-product}) as the \textit{PI matrix algebra} $\mathcal{M}_{AB}$. A subspace of $\mathcal{M}_{AB}$ that is closed under multiplication is called the \textit{PI matrix subalgebra} $\mathcal{M}'_{AB}$.

\begin{figure}
\includegraphics[width=3.4in]{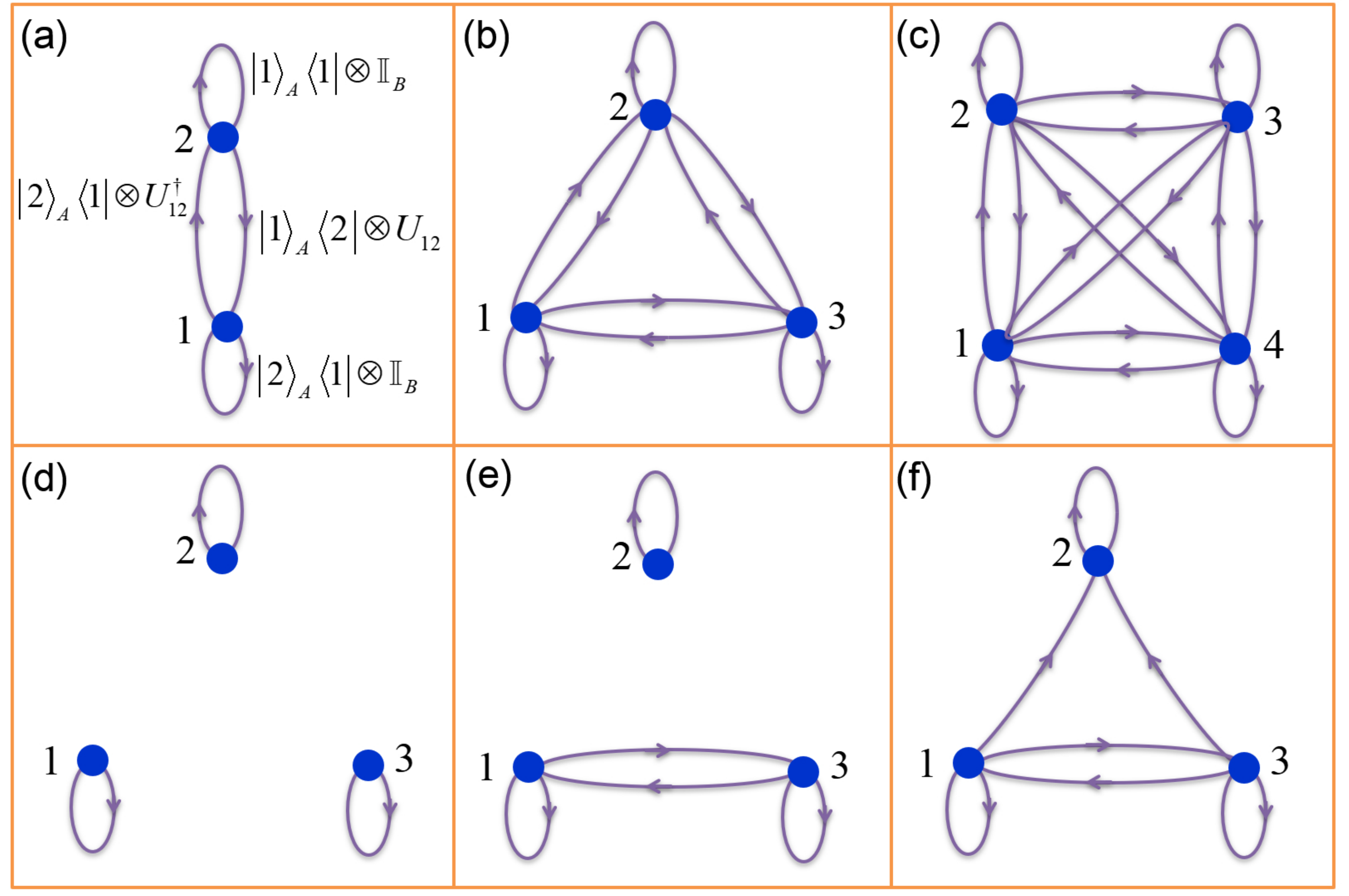}
\caption{(a-c) Diagrams of the basis of PI matrix algebras for $d_A=2,3,4$. (d-f) Diagrams of the basis of some PI matrix subalgebras for $d_A=3$. In the diagrams, the $d_A$ blue dots represent the ancilla states $\{|m\rangle_A\}_{m=1}^{d_A}$, and the loop or line with arrow pointing from $|n\rangle_A$ to $|m\rangle_A$ represent the base vector $|m\rangle_A\langle n|\otimes U_{mn}$, satisfying $U_{me}U_{en}=U_{mn}$ for $m,e,n\in[1,d_A]$. The number of base vectors is $d_A^2$ for a PI matrix algebra and smaller than $d_A^2$ for a PI matrix subalgebra. }
\label{basis}
\end{figure}

One can see that $\mathcal{M}_{AB}$ is isomorphic to $\mathcal{M}_{\rm A}$ (Table \ref{table}), since the multiplication operation is preserved by the map between them (see Appendix \ref{SI-A1} for the explicit form of the map). The bases of PI matrix algebras and subalgebras are represented diagrammatically in Fig. \ref{basis}. Note that the set of ancilla projection operators $\{|m\rangle_A\langle m|\}_{m=1}^{d_A}$ belong to $\mathcal{B}_{\rm A}$ and $\{|m\rangle_A\langle m|\otimes \mathbb{I}_B\}_{m=1}^{d_A}$ belong to $\mathcal{B}_{AB}$. Moreover, the PI matrix algebra is a self-ajoint algebra (closed under Hermitian conjugation) [Fig. \ref{basis}(a)-(c)], since $|m\rangle_A\langle n|\otimes U_{mn}=(|n\rangle_A\langle m|\otimes U_{nm})^{\dag}$. The PI matrix subalgebras can be self-adjoint [Fig. \ref{basis}(d) and \ref{basis}(e)] or non-self-adjoint [Fig. \ref{basis}(f)], but a non-self-adjoint PI matrix subalgebra can be directly extended to become self-adjoint. 

A remarkable feature of the matrix algebras is the path-independence property for a sequential product of its basis vectors. Consider the product of a sequence of elements in $\mathcal{B}_{\rm A}$, $|r\rangle_A\langle a|a\rangle_A\langle b|b\rangle_A\langle c|\cdots |e\rangle_A\langle i|=|r\rangle_A\langle i|$, with $i,a,b,c,\cdots,e,r\in[1,d_A]$.
%\begin{align}\label{}
%    |r\rangle_A\langle a|a\rangle_A\langle b|b\rangle_A\langle c|\cdots |e\rangle_A\langle i|=|r\rangle_A\langle i|,
%\end{align}
Such a product is determined only by the bra $_A\langle i|$ of the first element and the ket $|r\rangle_A$ of the final element, but independent of any other intermediate elements.
Likewise, a corresponding product of elements in $\mathcal{B}_{AB}$ is
\begin{align}\label{}
    &(|r\rangle_A\langle a|\otimes U_{ra})\cdot(|a\rangle_A\langle b|\otimes U_{ab})\cdot(|b\rangle_A\langle c|\otimes U_{bc}) \nonumber \\
    &\cdots (|e\rangle_A\langle i|\otimes U_{ei})=|r\rangle_A\langle i|\otimes U_{ri}.
\end{align}
For the diagrams in Fig. \ref{basis}, this means that any base vector $|r\rangle_A\langle i|\otimes U_{ri}$ depends only on the initial ancilla state $|i\rangle_A$ and final ancilla state $|r\rangle_A$, but independent of the detailed paths from $|i\rangle_A$ to $|r\rangle_A$. Then suppose that we make a preselection $|i\rangle_A$ and a postselection $|r\rangle_A$ of the ancilla states, the central system undergoes a deterministic unitary evolution $U_{ri}$ independent of any intermediate paths (corresponding to possible ancilla errors). Such a path-independence property of matrix algebras is the underlying principle for the PI gates.

\newcommand{\tabincell}[2]{\begin{tabular}{@{}#1@{}}#2\end{tabular}}
\begin{table*}[]
\begin{spacing}{1.6}
%\begin{tabular*}{16.5cm}{ccc}
\begin{tabular}{p{1.0cm} p{2.7cm} p{3.7cm} p{4.1cm} p{6.2cm}}
\hline
\hline
\multirow{2}{*}
{} & \tabincell{c}{$\mathcal{M}_{A}$}  & \tabincell{c}{$\mathcal{M}_{AB}$} &  \tabincell{c}{$\hat{\mathcal{M}}_{A}$}& \tabincell{c}{$\hat{\mathcal{M}}_{AB}$} \\
\hline
\tabincell{c}{Basis} & $\{|m\rangle_{A}\langle n|\}_{m,n=1}^{d_A}$ & $\{|m\rangle_{A}\langle n|\otimes U_{mn}\}_{m,n=1}^{d_A}$ &  $\{|mn\rangle\rangle_A\langle\langle pq|\}_{m,n,p,q=1}^{d_A}$  &  $\{|mn\rangle\rangle_A\langle\langle pq|\otimes (U_{mp}\otimes U^{*}_{nq})\}_{m,n,p,q=1}^{d_A}$  \\
\hline
\hline
\end{tabular}
\end{spacing}
\caption{ Different kinds of matrix algebras and their bases. ${M}_{A}$, $\mathcal{M}_{AB}$, $\hat{\mathcal{M}}_{A}$ and $\hat{\mathcal{M}}_{AB}$ are defined on the Hilbert space of the ancilla system,  the Hilbert space of the composite system, the HS space of the ancilla system and the HS space of the composite system, correspondingly. Note the isomorphism between $\mathcal{M}_{A}$ and $\mathcal{M}_{AB}$, and between $\hat{\mathcal{M}}_{A}$ and $\hat{\mathcal{M}}_{AB}$.
}
\label{table}
%\end{ruledtabular}
\end{table*}

\textit{Eigenvalues and eigenvectors of PI operators}.---Due to the isomorphism between $\mathcal{M}_{\rm A}$ and $\mathcal{M}_{AB}$, for any ancilla operator $H_{A}=\sum_{m,n}h_{mn}|m\rangle_A\langle n|$, we can define a corresponding PI operator in the composite system as
\begin{align}\label{}
    H_{AB}=\sum_{m,n}h_{mn}|m\rangle_A\langle n|\otimes U_{mn}.
\end{align}
The eigenvalues and eigenvectors of $H_{A}$ and $H_{AB}$ are related in the following way (see Appendix \ref{SI-A1} for the proof).

\textbf{Lemma 1}. Let $\{\lambda_i\}_{i=1}^{d_A}$ be the set of eigenvalues of $H_{A}$, the eigenvalues of $H_{AB}$ are still $\{\lambda_i\}_{i=1}^{d_A}$ but each with an algebraic multiplicity $d_B$. For each eigenvector  $|v\rangle_A=\sum_{m=1}^{d_A}c_{m}|m\rangle_A$ of $H_{A}$ with eigenvalue $\lambda$, $H_{AB}$ has a corresponding $d_B$-dimensional degenerate eigenspace spanned by $\{|v_{j}\rangle\}_{j=1}^{d_B}$ with $|v_{j}\rangle=\sum_{m=1}^{d_A}c_{m}|m\rangle_A\otimes U_{mk}|j\rangle_B$ with $k$ being an arbitrary integer within $[1,d_A]$.

%The operators in the PI matrix algebra have degenerate eigenspaces with the degeneracies being exactly the Hilbert space dimension of the central system.

\textit{PI propagators}.---The isomorphism between $\mathcal{M}_{A}$ and $\mathcal{M}_{AB}$ also facilitates the solution of the propagator generated by a PI operator $H_{AB}$, which we call the PI propagator. Suppose that $W_{A}(t,0)=e^{-i H_{A}t}=\sum_{m,n}\xi_{mn}(t)|m\rangle_{A}\langle n|$, then we have
\begin{align}\label{Wab}
    W_{AB}(t,0)=e^{-i H_{AB}t}=\sum_{m,n}\xi_{mn}(t)|m\rangle_{A}\langle n|\otimes U_{mn}.
\end{align}
Note that the analog still holds for time-dependent operators $H_{A}(t)$ and $H_{AB}(t)$. Moreover, we can easily prove the lemma below  (see Appendix \ref{SI-A2} for the proof).

\textbf{Lemma 2}. The propagator $W_{AB}(t,0)$ is in the PI matrix algebra $\mathcal{M}_{AB}$ if and only if its generator $H_{AB}\in\mathcal{M}_{AB}$.

The PI propagator has the special property
\begin{align}\label{WabP}
    P_{r}W_{AB}(t,0)P_{i}\propto |r\rangle_A\langle i|\otimes U_{ri},
\end{align}
where $P_{i}=|i\rangle\langle i|\otimes \mathbb{I}_B$ is an ancilla projection operator. So the PI propagation of the composite system with a preselection and postselection on the ancilla states results in a deterministic unitary operation on the central system. This can also be understood by the spectral properties of the PI operators in Lemma 1. $W_{AB}$ make the composite system evolve in different degenerate eigenspaces of $H_{AB}$, in exactly the same way as $W_A$ drives the ancilla system. The initial and final projections induce the transition
\begin{align}\label{}
    |i\rangle_A\otimes U_{ik}|\psi\rangle_B\rightarrow |r\rangle_A\otimes U_{rk}|\psi\rangle_B,
\end{align}
with the accompanying unitary evolution $U_{rk}U_{ik}^{\dagger}=U_{ri}$ on the central system, where $|\psi\rangle_B$ is an arbitrary central system state. Moreover, since this property holds true for any initial and final ancilla states, we can perform conditional operation until the gate succeeds.

Note that $W_{AB}$ represents a general class of PI propagators. Suppose that $W_{AB}(t,0)$ is the propagator in the interaction picture associated with an arbitrary diagonal Hamiltonian in the ancilla basis $H_0(t)=\sum_{m=1}^{d_A}|m\rangle_A\langle m|\otimes H_m(t)$ \cite{footnote2}, where $\{H_{m}(t)\}$ is a set of arbitrary time-dependent Hamiltonians on the central system. In the Schr\"{o}dinger's picture the propagator is
$W_{AB}^{(S)}(t,0)=\sum_{m,n}\xi_{mn}(t)|m\rangle_{A}\langle n|\otimes R_m(t)U_{mn}$, where $R_m(t)=\mathcal{T}\{e^{-i\int_{0}^{t}H_m(t')dt'}\}$ with $\mathcal{T}$ being the time-ordering operator. Then $P_{r}W_{AB}^{(S)}(t,0)P_{i}\propto |r\rangle_A\langle i|\otimes R_r(t)U_{ri}$, so the central system still undergoes a unitary evolution.

\textit{PI matrix algebra for the Hilbert-Schmidt space}.---The PI matrix algebra can be directly extended to the HS space. For the ancilla system, its HS space has an orthonormal basis $\{|mn\rangle\rangle_A\}_{m,n=1}^{d_A}$, where $|mn\rangle\rangle_A=|m\rangle_A\langle n|$, while the operators in the HS space lie in the matrix algebra $\hat{\mathcal{M}}_A$ spanned by the basis $\hat{\mathcal{B}}_A=\{|mn\rangle\rangle_A\langle\langle pq|\}_{mn,pq}$ \cite{Ernst1987, Albert2014}. A general superoperator for the ancilla is $\mathcal{H}_A=\sum_{mn,pq}h_{mn,pq}|m\rangle_A\langle p|(\cdot)|q\rangle_A\langle n|$ with $(\cdot)$ denoting an arbitrary ancilla operator, corresponding to the operator $\hat{\mathcal{H}}_A=\sum_{mn,pq}h_{mn,pq}|mn\rangle\rangle_A\langle\langle pq|$ in the HS space. For example, $X_A(\cdot)Y_A\leftrightarrow \sum_{mn,pq}x_{mp}y_{qn}|mn\rangle\rangle_A\langle\langle pq|=X_A\otimes Y_A^T$, where $X_A, Y_A\in \mathcal{M}_A$ and $Z_A^T$ is the transpose of $Z_A$.

For the composite system, we can formulate a general class of superoperators by restricting the left and right multiplication operators to vectors in the PI matrix algebra $\mathcal{M}_{AB}$,
\begin{align}\label{Sab}
    &\mathcal{H}_{AB}=\sum_{mn,pq}h_{mn,pq}(|m\rangle_A\langle p|\otimes U_{mp})(\cdot)(|q\rangle_A\langle n|\otimes U_{nq}^{\dagger}), \nonumber \\
    &\Updownarrow \nonumber \\
    &\hat{\mathcal{H}}_{AB}=\sum_{mn,pq}h_{mn,pq}|mn\rangle\rangle_A\langle\langle pq|\otimes (U_{mp}\otimes U_{nq}^{*}),
\end{align}
where $(\cdot)$ denotes an arbitrary operator of the composite system and $U_{nq}^{*}$ is the complex conjugate of $U_{nq}$. With $X_{AB}, Y_{AB}\in \mathcal{M}_{AB}$, $X_{AB}(\cdot)Y_{AB}\leftrightarrow \sum_{mn,pq}x_{mp}y_{qn}|mn\rangle\rangle_A\langle\langle pq|\otimes (U_{mp}\otimes U^{*}_{nq})$. This motivates the following definition.

\textbf{Definition 2}: \textbf{PI matrix algebra for HS space}. Consider a vector space with the basis $\hat{\mathcal{B}}_{AB}=\{|mn\rangle\rangle_A\langle\langle pq|\otimes (U_{mp}\otimes U^{*}_{nq})\}_{m,n,p,q=1}^{d_A}$ over $\mathbb{C}$, where $\{U_{mn}\}$ is the same set of unitary operators on the central system as that in Definition 1. Then a multiplication operation can be defined as
\begin{align}\label{def-product-HS}
    &\left[|mn\rangle\rangle_A\langle\langle pq|\otimes (U_{mp}\otimes U^{*}_{nq})\right]\left[|rs\rangle\rangle_A\langle\langle tv|\otimes (U_{rt}\otimes U^{*}_{sv})\right] \nonumber \\
    &=\delta_{pr}\delta_{qs}|mn\rangle\rangle_A\langle\langle tv|\otimes (U_{mt}\otimes U^{*}_{nv}),
\end{align}
with $m,n,p,q,r,s,t,v\in[1,d_A]$. Define this vector space with the multiplication operation in Eq. (\ref{def-product-HS}) as the \textit{PI matrix algebra} $\hat{\mathcal{M}}_{AB}$  \textit{for the HS space}.% A subspace of $\mathcal{M}_{AB}$ that is closed under multiplication is called the \textit{PI matrix subalgebra} $\mathcal{N}'_{AB}$.

Then any superoperator $\hat{\mathcal{H}}_{AB}$ [Eq. (\ref{Sab})] is a vector in a PI matrix algebra $\hat{\mathcal{M}}_{AB}$. The spectral properties of $\hat{\mathcal{H}}_{AB}$ can be determined in analog to Lemma 1, except that the algebraic multiplicity of eigenvalues becomes $d_B^2$  (see Appendix \ref{SI-A3}). Moreover, according to Lemma 2, the propagator $\hat{\mathcal{W}}_{AB}(t,0)=e^{-i\hat{\mathcal{H}}_{AB}t}$ is still in $\hat{\mathcal{M}}_{AB}$. Such a propagator has the same PI property as that in Eq. (\ref{WabP})
\begin{align}\label{WabPh}
    \hat{\mathcal{P}}_r \hat{\mathcal{W}}_{AB}(t,0)\hat{\mathcal{P}}_i\propto |rr\rangle\rangle_A\langle\langle ii|\otimes (U_{ri}\otimes U_{ri}^{*}),
\end{align}
where $\hat{\mathcal{P}}_{i}=|ii\rangle\rangle_A\langle\langle ii|\otimes \hat{\mathbb{\mathcal{I}}}_B$ is the superoperator for the ancilla projection with $\hat{\mathbb{\mathcal{I}}}_B$ being the identity operation in the HS space of the central system. We term this condition the \textit{PI gate condition}. For the closed-system evolution of the composite system driven by a single Hamiltonian, Eq. (\ref{WabPh}) is equivalent to Eq. (\ref{WabP}). However, the power of Eq. (\ref{WabPh}) shows up when dealing the open-system evolution of the composite system, where we can treat the Hamiltonian and the dissipation operators on the same footing.

\textit{Path-independence for ancilla noise}.---Suppose the ancilla suffers from Markovian noise and the dynamics of the composite system is described by
\begin{eqnarray}\label{Mar}
    \frac{d{\rho}}{dt}=\mathcal{L}_{AB}(\rho)=i[\rho, H_{AB}] +\sum_i\mathcal{D}[K_{i}]\rho,
\end{eqnarray}
where $D[K]\rho=K\rho K^{\dagger}-\{K^{\dagger}K,\rho\}/2$ is the Lindbladian dissipator. In the HS space, the Liouville superoperator $\mathcal{L}_{AB}$ becomes
\begin{align}\label{Lab}
    &\hat{\mathcal{L}}_{AB}=-i\hat{\mathcal{H}}_{AB}=-i(H_{AB}\otimes \mathbb{I}_{AB}-\mathbb{I}_{AB}\otimes H_{AB}^{*}) \nonumber \\
    &+\frac{1}{2}\sum_i[2K_{i}\otimes K_{i}^{*}-K_{i}^{\dag}K_{i}\otimes \mathbb{I}_{AB}-\mathbb{I}_{AB}\otimes (K_{i}^{\dag}K_{i})^{*}].
\end{align}

The criteria of PI gates in \cite{Ma2020} relies on a Dyson expansion of the propagator $\hat{\mathcal{W}}_{AB}(t,0)=e^{\hat{\mathcal{L}}_{AB}t}$ generated by the Liouville superoperator. With the PI matrix algebra for the HS space, we can provide an exact and unifying criteria for PI gates.

\textbf{Theorem 1}. The PI gate condition [Eq. (\ref{WabPh})] is exactly satisfied if $H_{AB}$ and $K_i$ for all $i$ are in the same PI matrix algebra $\mathcal{M}_{AB}$ or subalgebra $\mathcal{M}'_{AB}$.
\begin{proof}
---If $H_{AB}$, $K_i$ $\in \mathcal{M}_{AB}$, then $K_i^{\dagger}\in \mathcal{M}_{AB}$ since $\mathcal{M}_{AB}$ is self-adjoint or closed under Hermitian conjugation, so $K_i^{\dagger}K_i\in \mathcal{M}_{AB}$. Then from Eq. (\ref{Sab}) we obtain $\hat{\mathcal{L}}_{AB}\in \hat{\mathcal{M}}_{AB}$. According to Lemma 2, $\hat{W}(t,0)\in \hat{\mathcal{M}}_{AB}$, so the PI gate condition is satisfied. If $H_{AB}$, $K_i$ $\in \mathcal{M}'_{AB}$, we can first extend $\mathcal{M}'_{AB}$ to be self-adjoint if it is not, then similar conclusion can be reached as in the former case.
\end{proof}

As a special case of of PI gates, error-transparent gates for QEC codes have been theoretically proposed \cite{Vy2013,Kapit2018} and recently experimentally demonstrated against a specific system error \cite{Ma2020}. The error transparency requires the physical Hamiltonian commutes with the errors when acting on the QEC code subspace (or the commutators of the physical Hamiltonian and errors are proportional to the errors). By relating the error syndromes of a QEC code with the ancilla states in PI gates, we show the error-transparency condition can be interpreted as a special case of Theorem 1 (see Appendix \ref{SI-C1}).

Theorem 1 also unifies the path-independence criteria for the ancilla dephasing and relaxation errors \cite{Ma2020} (see Appendix \ref{SI-C2}), experimentally relevant to the PI SNAP gates in superconducting circuits \cite{Reinhold2020}. Since the ancilla dephasing operator $K_i\propto \sum_{m=1}^{d_A}\Delta_{im}|m\rangle_A\langle m|\otimes \mathbb{I}_B$ is automatically in any PI matrix algebra, so the PI gate condition is exactly satisfied with a PI control Hamiltonian $H_{AB}$. However, the ancilla relaxation operator $K_j\propto |m\rangle_A\langle n|\otimes \mathbb{I}_B$ requires additional conditions to lie in the same PI matrix algebra with $H_{AB}$, which has been analyzed in \cite{Ma2020} (also see Appendix \ref{SI-C}). Moreover, if the condition of theorem 1 is not exactly satisfied, it is still possible to have an approximate PI condition up to the leading-order Dyson expansions of the Liouville superoperator \cite{Ma2020} (see Appendix \ref{SI-B}).

\textit{Summary}.---With the discovery of PI matrix algebra, we reveal the elegant mathematical structure of PI quantum control. We also find that the PI matrix algebras can not only be formed by operators in the Hilbert space of a composite system, but also by the operators in the HS space of the same system. This permits us to treat the open system dynamics of the composite system in a rigorous way and provide an exact and unifying criteria for PI gates. The PI matrix algebra is also interesting fundamentally, since the PI operators have peculiar spectral properties and degenerate eigenspaces (see Lemma 1). Moreover, PI operators is generally non-Hermitian, so it will be interesting to study the rich non-Hermitian properties of PI operators \cite{Ashida2020}, such as pseduo-Hermiticity and exceptional points, and expore their physical implications.

\begin{acknowledgments}
We thank Mengzhen Zhang for helpful discussions. We acknowledge support from the ARO (W911NF-18-1-0020, W911NF-18-1-0212), ARO MURI (W911NF-16-1-0349, W911NF-21-1-0325), AFOSR MURI (FA9550-19-1-0399, FA9550-21-1-0209), NSF (EFMA-1640959, OMA-1936118, EEC-1941583, OMA-2137642), NTT Research, and the Packard Foundation (2020-71479). W. -L. Ma also acknowledges support from the Startup Foundation of Institute of Semiconductors, Chinese Academy of Sciences (No. E0SEBB11), and National Natural Science Foundation of China (Grant No. 12174379).
\end{acknowledgments}

\appendix
\section{Proofs of Lemmas in the main text}
\subsection{Proof of Lemma 1}\label{SI-A1}
Due to the isomorphism between the ordinary matrix algebra $\mathcal{M}_{\rm A}$ and the PI matrix algebra $\mathcal{M}_{AB}$, for any operator of the ancilla in $\mathcal{M}_{\rm A}$, we can define a corresponding operator of both the ancilla and central systems in $\mathcal{M}_{AB}$ as
\begin{subequations}\label{PI-H}
\begin{align}\label{}
    H_{A}&=\sum_{m,n}h_{mn}|m\rangle_A\langle n|,  \\
    &\Updownarrow \nonumber \\
    H_{AB}&=\sum_{m,n}h_{mn}|m\rangle_A\langle n|\otimes U_{mn}.
\end{align}
\end{subequations}
We can explicitly construct the map as a unitary transformation
\begin{align}\label{Uab}
    U_{AB}(H_A\otimes \mathbb{I}_B)U_{AB}^{\dagger}=H_{AB},
\end{align}
with
\begin{align}\label{}
    U_{AB}=\sum_{m=1}^{d_A} |m\rangle_A\langle m|\otimes U_{mk},
\end{align}
being a unitary matrix implementing different unitary operations on the central systems dependent on the ancilla states and $k$ being an arbitrary integer within $[1,d_A]$.

The eigenvalues and eigenvectors of $H_{A}$ and $H_{AB}$ are closely related in the following way.

\textbf{Lemma 1}. Let $\{\lambda_i\}_{i=1}^{d_A}$ be the set of eigenvalues of $H_{A}$, the eigenvalues of $H_{AB}$ are still $\{\lambda_i\}_{i=1}^{d_A}$ but each with an algebraic multiplicity $d_B$. For each eigenvector  $|v\rangle_A=\sum_{m=1}^{d_A}c_{m}|m\rangle_A$ of $H_{A}$ with eigenvalue $\lambda$, $H_{AB}$ has a corresponding $d_B$-dimensional degenerate eigenspace spanned by $\{|v_{j}\rangle\}_{j=1}^{d_B}$ with $|v_{j}\rangle=\sum_{m=1}^{d_A}c_{m}|m\rangle_A\otimes U_{mk}|j\rangle_B$ with $k$ being an arbitrary integer within $[1,d_A]$.

\begin{proof}
---The eigenvalues of $H_{AB}$ can be obtained by two different approaches. In the first approach, notice that $H_A\otimes \mathbb{I}_B$ and $H_{AB}$ are related through a unitary transformation [Eq. (\ref{Uab})], so they must have the same set of eigenvalues. For the ancilla system alone, denote the eigenvalues of $H_{A}$ as $\{\lambda_i\}_{i=1}^{d_A}$. For the composite system, the eigenvalues of $H_{A}\otimes \mathbb{I}_B$ are still $\{\lambda_i\}_{i=1}^{d_A}$ but each with an algebraic multiplicity $d_B$. So is the case with $H_{AB}$.

In the second approach, we try to calculate the determinant associated with $H_{AB}$. Since $\{\lambda_i\}_{i=1}^{d_A}$ are the set of eigenvalues of $H_{A}$, then we have
\begin{align}\label{}
    {\rm det}(H_A-\lambda \mathbb{I}_{A})=\prod_{i=1}^{d_A}(\lambda-\lambda_i).
\end{align}
The determinant can be obtained by successively performing the 2-partition of a matrix. For a $d\times d$ matrix $H$, denote $\alpha\subset\{1,\cdots,d\}$ as an index set and $\alpha^c\subset\{1,\cdots,d\}\setminus\alpha$ as the index set complementary to $\alpha$. For index sets $\alpha, \beta$, denote by $H[\alpha, \beta]$ the submatrix of entries that lie in rows of $H$ indexed by $\alpha$ and the columns indexed by $\beta$, and simplify $H[\alpha, \alpha]$ as $H[\alpha]$. Then for a nonsingular $H[\alpha]$, we have
\begin{align}\label{}
    &{\rm det} (H) \nonumber \\
    &={\rm det}\left(H[\alpha]\right){\rm det}\left(H[\alpha^c]-H[\alpha^c, \alpha]H[\alpha]^{-1}H[\alpha, \alpha^c]\right),
\end{align}
where $H[\alpha^c]-H[\alpha^c, \alpha]H[\alpha]^{-1}H[\alpha, \alpha^c]$ is the Schur complement of $H[\alpha]$ \cite{Horn2013}.
To compute ${\rm det}(H_A-\lambda \mathbb{I}_{A})$, we can first perform the 2-partition of $H_{A}$ with $\alpha=\{1\}$, then similarly perform the 2-partition of the remaining Schur complement. The process is repeated until the final Schur complement is a single entry, so that the determinant can be expressed as sequential product.
To compute ${\rm det}(H_{AB}-\lambda \mathbb{I}_{AB})$, we can similarly perform the successive 2-partition with respect to the ancilla state index only. Since $U_{mm}=\mathbb{I}_B$, one can verify that
\begin{align}\label{}
    {\rm det}(H_{AB}-\lambda \mathbb{I}_{AB})=\left[{\rm det}(H_A-\lambda \mathbb{I}_{A})\right]^{d_B}=\prod_{i=1}^{d_A}(\lambda-\lambda_i)^{d_B}.
\end{align}
Therefore, $H_{AB}$ has the same set of eigenvalues as that of $H_{A}$ but each with an algebraic multiplicity $d_B$.

From  Eq. (\ref{Uab}), the eigenvector of $H_{AB}$ can be obtained directly from that of $H_A\otimes\mathbb{I}_B$ as
\begin{align}\label{}
    |v_{j}\rangle=U_{AB}|v\rangle_A|j\rangle_B=\sum_{n=1}^{d_A}c_{n}|m\rangle_A\otimes U_{nk}|j\rangle_B.
\end{align}
We can easily verify that $|v_{j}\rangle$ is an eigenvector of $H_{AB}$. Since $H_{A}|v\rangle_A=\lambda |v\rangle_A$, we have $\sum_{m=1}^{d_A}h_{nm}c_{m}=\lambda c_{n}$, then
\begin{align}\label{}
    H_{AB}|v_{j}\rangle&=\sum_{m,n=1}^{d_A}h_{nm}c_{m}|n\rangle_A\otimes U_{nm}U_{mk}|j\rangle_B \nonumber \\
    &=\lambda\sum_{n=1}^{d_A}c_{n}|m\rangle_A\otimes U_{nk}|j\rangle_B \nonumber \\
    &=\lambda|v_{j}\rangle,
\end{align}
 where we use $U_{nm}U_{mk}=U_{nk}$.

\end{proof}

\subsection{Proof of Lemma 2}\label{SI-A2}
\textbf{Lemma 2}. The propagator $W_{AB}(t,0)$ is in the PI matrix algebra $\mathcal{M}_{AB}$ if and only if its generator $H_{AB}\in\mathcal{M}_{AB}$.
\begin{proof}
---If $H_{AB}\in\mathcal{M}_{AB}$, then $H_{AB}^n\in\mathcal{M}_{AB}$ for any positive integer $n$, then $W_{AB}(t,0)=e^{-iH_{AB}t}=\sum_{n=0}^{\infty }\frac{(-i)^n}{n!}H_{AB}^n\in\mathcal{M}_{AB}$ (the above series always converges for any matrix operator $H_{AB}$ in a finite-dimensional vector space).

 Conversely, if $W_{AB}(t,0)\in\mathcal{M}_{AB}$ for any $t\in[0,\infty)$, then $\frac{dW_{AB}(t,0)}{dt}\in\mathcal{M}_{AB}$. So $H_{AB}=i\frac{dW_{AB}(t,0)}{dt}|_{t=0}\in\mathcal{M}_{AB}$. The proof can be easily generalized to the case of a time-dependent $H_{AB}(t)$.

\end{proof}

\subsection{Eigenvalues and eigenvectors of PI operators in the HS space of the composite system}\label{SI-A3}
For the HS space, there also exists the isomorphism between $\hat{\mathcal{M}}_{\rm A}$ and $\hat{\mathcal{M}}_{AB}$, so we have the following correspondence,
\begin{subequations}\label{}
\begin{align}\label{}
    \hat{\mathcal{H}}_{A}&=\sum_{mn,pq}h_{mn,pq}|mn\rangle\rangle\langle\langle pq|,  \\
    &\Updownarrow \nonumber \\
    \hat{\mathcal{H}}_{AB}&=\sum_{mn,pq}h_{mn,pq}|mn\rangle\rangle\langle\langle pq|\otimes (U_{mp}\otimes U_{nq}^{*}).
\end{align}
\end{subequations}
Then from Lemma 1, we can directly deduce that the eigenvalues and eigenvectors of $\hat{\mathcal{H}}_{A}$ and $\hat{\mathcal{H}}_{AB}$ are related in the following way.

\textbf{Lemma 3}. Let $\{\lambda_i\}_{i=1}^{d_A^2}$ be the set of eigenvalues of $\hat{\mathcal{H}}_{A}$, the eigenvalues of $\hat{\mathcal{H}}_{AB}$ are $\{\lambda_i\}_{i=1}^{d_A^2}$ but each with an algebraic multiplicity $d_B^2$. For each eigenvector  $|v\rangle\rangle_A=\sum_{m,n=1}^{d_A}c_{mn}|mn\rangle\rangle_A$ of $\hat{\mathcal{H}}_{A}$ with eigenvalue $\lambda$, $\hat{\mathcal{H}}_{AB}$ has a corresponding $d_B^2$-dimensional degenerate eigenspace spanned by $\{|v_{jk}\rangle\rangle\}_{j,k=1}^{d_B}$ with $|v_{jk}\rangle\rangle=\sum_{m,n=1}^{d_A}c_{mn}|mn\rangle\rangle_A\otimes (U_{ml}\otimes U^{*}_{nl})|jk\rangle\rangle_B$ with $l$ being an arbitrary integer within $[1,d_A]$.

\section{Approximate PI gate condition}\label{SI-B}
If a general superoperator $\hat{\mathcal{H}}_{AB}$ can be divided into two parts as
\begin{align}\label{}
    \hat{\mathcal{H}}_{AB}=\hat{\mathcal{H}}_{\rm eff}+\hat{\mathcal{V}},
\end{align}
where $\hat{\mathcal{H}}_{\rm eff}$ is the dominant part and $\hat{\mathcal{V}}$ is a perturbation, then the open-system evolution driven by $\hat{\mathcal{H}}_{AB}$ can be represented by a generalized Dyson expansion as
\begin{align}\label{Wabh}
    \hat{\mathcal{W}}_{AB}(t,0)=e^{-i(\hat{\mathcal{H}}_{\rm eff}+\hat{\mathcal{V}})t}=\sum_{p=0}^{\infty}\hat{\mathcal{W}}_p(t,0),
\end{align}
with
\begin{eqnarray}
    \hat{\mathcal{W}}_0(t,0)=e^{-i\hat{\mathcal{H}}_{\rm eff}t},
\end{eqnarray}
\begin{equation}\label{}
\begin{split}
 &\hat{\mathcal{W}}_p(t,0)={\int_0^t {d{t_p}} } \cdots \int_0^{{t_3}} {d{t_{2}}} \int_0^{{t_2}} {d{t_1}}
  \hat{\mathcal{W}}_0\left( {t,{t_p}} \right) \\
 &\times\hat{\mathcal{V}}\cdots
 \hat{\mathcal{V}}\hat{\mathcal{W}}_0\left( {{t_2},{t_{1}}} \right)\hat{\mathcal{V}}\hat{ \mathcal{W}}_0\left( {{t_1},0}\right), ~~p\geq1,
\end{split}
\end{equation}
where $\hat{\mathcal{W}}_0(t_2,t_1)=e^{-i\hat{\mathcal{H}}_{\rm eff}(t_2-t_1)}$.

In cases where the exact PI gate condition [Eq. (\ref{WabPh})] cannot be satisfied, we can formulate the approximate PI gate condition as follows \cite{Ma2020}. Suppose that
\begin{align}\label{PI-appro}
    \hat{\mathcal{P}}_r \left[\sum_{p=0}^{k}\hat{\mathcal{W}}_p(t,0)\right]\hat{\mathcal{P}}_i\propto |rr\rangle\rangle_A\langle\langle ii|\otimes (U_{ri}\otimes U_{ri}^{*}),
\end{align}
applies for $k\leq n$ but does not hold for $k>n$, then we say the PI gate condition is satisfied up to the $n$th order from $|i\rangle_A$ to $|r\rangle_A$.

Just as $W_{AB}(t,0)$ [Eq. (\ref{Wab})] represents a general class of PI propagators in the Hilbert space of the composite system, $\hat{\mathcal{W}}_{AB}(t,0)$ [Eq. (\ref{Wabh})] represents a general class of PI propagators in the HS space of the composite system. Suppose $\hat{\mathcal{W}}_{AB}(t,0)$ is the propagator in the interaction picture associated with a Hermitian Hamiltonian $\hat{\mathcal{H}}_{0}(t)=H_0(t)\otimes \mathbb{I}_{AB}+\mathbb{I}_{AB}\otimes H_0^*(t)$, where $H_0(t)=\sum_{m}|m\rangle\langle m|\otimes H_m(t)$. Then in the Schr\"{o}dinger's picture, the Hamiltonian and propagator in the HS space are
\begin{align}\label{}
    &\hat{\mathcal{H}}_{AB}^{(S)}(t)=\hat{\mathcal{R}}(t)\hat{\mathcal{H}}_{AB}\hat{\mathcal{R}}^{\dagger}(t)+i\frac{\partial \hat{\mathcal{R}}(t)}{\partial t}\hat{\mathcal{R}}^{\dagger}(t), \\
    & \hat{\mathcal{W}}_{AB}^{(S)}(t,0)=\hat{\mathcal{R}}(t)\hat{\mathcal{W}}_{AB}(t,0),
\end{align}
with
\begin{align}\label{}
    &\hat{\mathcal{R}}(t)=e^{-i\hat{\mathcal{H}}_{0}t}=R(t)\otimes R^{*}(t) \nonumber \\
    &=\sum_m |mm\rangle\rangle_A\langle\langle mm|\otimes [R_m(t)\otimes R^{*}_m(t)],
\end{align}
where $R_m(t)=\mathcal{T}\{e^{-i\int_{0}^{t}H_m(t')dt'}\}$. Then the exact PI gate condition in the Schr\"{o}dinger's picture is
\begin{align}\label{}
    &\hat{\mathcal{P}}_r \hat{\mathcal{W}}_{AB}^{(S)}(t,0)\hat{\mathcal{P}}_i \nonumber \\
    &\propto |rr\rangle\rangle_A\langle\langle ii|\otimes [R_r(t)U_{ri}\otimes R^{*}_r(t)U_{ri}^{*}].
\end{align}
Moreover, since we can have a generalized Dyson expansion of $\hat{\mathcal{W}}_{AB}^{(S)}(t,0)$ as
\begin{align}\label{}
 \hat{\mathcal{W}}_{AB}^{(S)}(t,0)=\sum_{p=0}^{\infty}\hat{\mathcal{W}}_p^{(S)}(t,0)=\sum_{p=0}^{\infty}\hat{\mathcal{R}}(t)\hat{\mathcal{W}}_p(t,0),
\end{align}
the approximate PI gate condition in the Schr\"{o}dinger's picture is
\begin{align}\label{PI-appro-S}
    &\hat{\mathcal{P}}_r \left[\sum_{p=0}^{k}\hat{\mathcal{W}}^{(S)}_p(t,0)\right]\hat{\mathcal{P}}_i \nonumber \\
    &\propto |rr\rangle\rangle_A\langle\langle ii|\otimes [R_r(t)U_{ri}\otimes R_r^*(t)U_{ri}^{*}].
\end{align}

With the exact PI gate condition satisfied, the central system undergoes a unitary evolution irrespective of the initial and final ancilla states (or the approximate gate condition is satisfied up to infinite-order for any initial and final ancilla states). But for approximate PI gates, the approximate PI condition is often satisfied up to different orders depending on the initial and final ancila states. We will give examples to illustrate this point in the next section (also see the Supplementary Information of \cite{Ma2020} for details).

The Liouville superoperator $\hat{\mathcal{L}}_{AB}$ [Eq. (\ref{Lab})] is often divided into two parts as
\begin{align}\label{}
    \hat{\mathcal{L}}_{AB}=-i\hat{\mathcal{H}}_{AB}=\hat{\mathcal{L}}_{\rm eff}+\hat{\mathcal{S}},
\end{align}
with
\begin{align}\label{}
    &\hat{\mathcal{L}}_{\rm eff}=-i(H_{\rm eff}\otimes \mathbb{I}_{AB}-\mathbb{I}_{AB}\otimes H_{\rm eff}^{*}) \\
    &\hat{\mathcal{S}}=\sum_iK_{i}\otimes K_{i}^{*},
\end{align}
with $H_{\rm eff}=H_{AB}-i\sum_i K_i^{\dagger}K_i/2$. Here $\hat{\mathcal{L}}_{\rm eff}$ generates the no-jump evolution with the non-Hermitian Hamiltonian $H_{\rm eff}$, while $\hat{\mathcal{S}}$ induces the quantum jumps during the no-jump evolution. The approximate PI condition [Eq. (\ref{PI-appro}) or Eq. (\ref{PI-appro-S})] for such a division of the Liouville superoperator is just the definition of path independence in \cite{Ma2020}.

\section{Examples of PI gates}\label{SI-C}
To illustrate how to use Theorem 1, we present two examples of PI gates, including error-transparent gates \cite{Vy2013,Kapit2018,Ma2020b} and PI SNAP gates \cite{Ma2020,Reinhold2020}. In the first example, we show that the error transparency condition can be reinterpreted as a special use of Theorem 1. In the second example, we show that the path independence criteria for ancilla dephasing and relaxation errors in PI SNAP gates can be unified by Theorem 1. For both examples, we also briefly discuss how the approximate gate condition can be applied.

\begin{figure}
\includegraphics[width=3.4in]{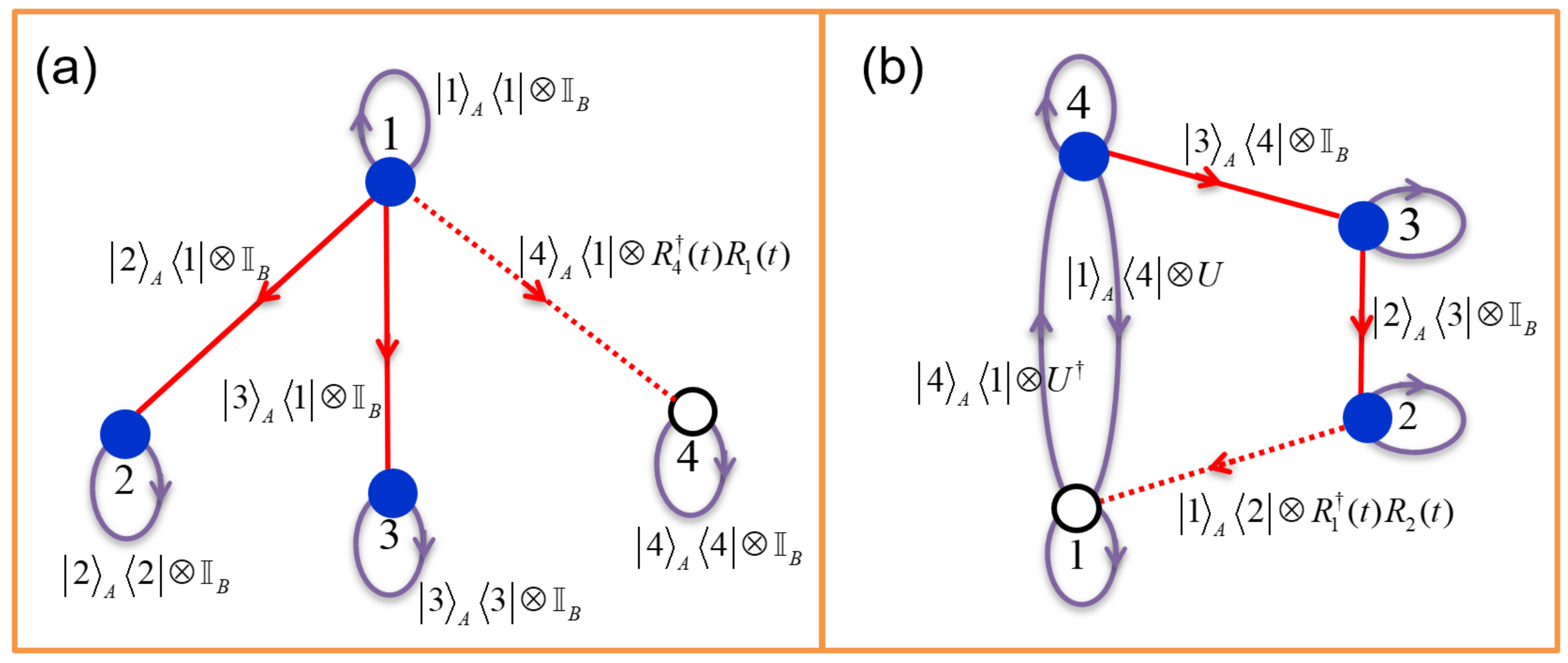}
\caption{Diagrams of the bases of PI matrix subalgebras with $d=4$ for (a) error-transparent gates and (b) PI SNAP gates. Here the purple loops or lines represent the base vectors for the no-jump evolution, and the red (red-dashed) lines represent the ancilla relaxation errors that can (cannot) form a PI matrix subalgebra with the no-jump evolution. }
\label{example}
\end{figure}

\subsection{Error-transparent gates}\label{SI-C1}
Below we present the model for error-transparent gates, by relating the ancilla states in this paper and the error syndromes of QEC codes. The Hilbert space of the central system with the ancilla in state $|1\rangle_A$ can be regarded as the logical subspace of a QEC code, while the Hilbert spaces of the central system with the ancilla in states $|2\rangle_A, \cdots, |d_A\rangle_A$ are the error subspaces. The correctable errors are $K_i=\sqrt{\gamma_i}|i\rangle_A\langle 1|\otimes \mathbb{I}_B$ with $i\in[2,d_A]$.

Consider the total Hamiltonian of the composite system as $H_{0}(t)=\sum_{m=1}^{d_A} |m\rangle_A \langle m|\otimes H_{m}(t)$, which may include both the static and control Hamiltonians. Such a Hamiltonian is error-transparent if
 \begin{align}\label{ET}
    [K_i,H_{0}(t)]=\lambda_{1i}K_i,
\end{align}
which is satisfied for any $i\in[2,d_A]$. One can see that this condition is equivalent to $H_1(t)-H_i(t)=\lambda_{1i}\in\mathbb{R}$, or $\{|m\rangle_A\}_{m=1}^{d_A}$ all belong to a noiseless ancilla subspace (NAS) defined in \cite{Ma2020}. With this condition, an error during the central (logical) system gate time is equivalent to an error after the gate (apart from a trivial phase factor), so the central system gate can be recovered by error correction of the ancilla system. Note that the error transparency condition in Eq. (\ref{ET}) is often more stringent than that in \cite{Kapit2018} (which only requires that Eq. (\ref{ET}) only satisfies when acting on the logical subspace).

We now show that the error transparency condition is equivalent to Theorem 1. We first move to the interaction picture associated with $H_0(t)$, so that the Hamiltonian vanishes. If Eq. (\ref{ET}) is satisfied, the error operator becomes
\begin{align}\label{}
    K_i(t)&=\sqrt{\gamma_i}|i\rangle_A\langle 1|\otimes R_i^{\dagger}(t)R_1(t) \nonumber \\
    &=\sqrt{\gamma_i}e^{-i\lambda_{1i}t}|i\rangle_A\langle 1|\otimes \mathbb{I}_B,
\end{align}
where $R_m(t)=\mathcal{T}\{e^{-i\int_{0}^{t}H_m(t')dt'}\}$. In this case, the set of error operators $\{K_i(t)\}$ definitely belong to an self-ajoint PI matrix subalgebra with the basis
\begin{align}\label{}
\{|m\rangle_A\langle m|\otimes \mathbb{I}_B\}_{m=1}^{d_A}\cup \{|i\rangle_A\langle 1|\otimes \mathbb{I}_B, |1\rangle_A\langle i|\otimes \mathbb{I}_B\}_{i=2}^{d_A}, \nonumber \\
\end{align}
so the PI gate condition is exactly satisfied for any initial and final ancilla states.

However, if some error operator $K_j$ does not satisfy Eq. (\ref{ET}) (i.e. $H_1(t)-H_j(t)$ is a non-trivial operator), the PI gate condition cannot be exactly satisfied for all initial and final states, but we can still use the approximate PI gate condition [Eq. (\ref{PI-appro})]. For the diagram in Fig. \ref{example}(a), we conclude that the PI gate condition is satisfied up to infinite-order (or exactly satisfied) from $|1\rangle_A$ to $|2\rangle_A$ or $|3\rangle_A$, but is satisfied only up to the zeroth order from $|1\rangle_A$ to $|3\rangle_A$.

\subsection{PI SNAP gates}\label{SI-C2}
The SNAP gate on the cavity, $S(\vec \varphi)=\sum_{n =0}^{\infty} e^{i\varphi_n}|n\rangle\langle n|$, imparts arbitrary phases $\vec \varphi=\{\varphi_n\}_{n=0}^{\infty}$ to the different Fock states of the cavity \cite{Krastanov2015,Heeres2015}. The SNAP gate, aided by $d_A$-level ancilla, can be made fault-tolerant against ancilla dephasing and relaxation errors with the PI design \cite{Ma2020,Reinhold2020}, therefore termed the PI SNAP gates. Below we present the model for PI SNAP gates.

Consider a static Hamiltonian $H_0=|1\rangle_A\langle 1|\otimes H_{1}+\sum_{m=2}^{d_A}|m\rangle_A\langle m|\otimes H_{2}$, where $H_{1}$, $H_{2}$ are Hamiltonians of the central system differing by some non-trivial operator (the constant ancilla state energy terms are neglected here). The control Hamiltonian is $H_{\rm c}(t)=\Omega(|1\rangle_A \langle d|\otimes R_1(t)UR_2^{\dagger}(t)+{\rm H.c.})$ with $U=S(\vec \varphi)$ and $R_j(t)=e^{-iH_jt} (j=1,2)$. One can see that the ancilla states $\{|m\rangle\}_{m=2}^{d_A}$ form a NAS.
The ancilla errors include the dephasing errors $\{\sqrt{\kappa_m}|m\rangle_A\langle m|\otimes \mathbb{I}_B\}_{m=1}^{d_A}$ and the relaxation errors $\{\sqrt{\gamma_m}|m-1\rangle_A\langle m|\otimes \mathbb{I}_B\}_{m=2}^{d_A}$.

To use Theorem 1, we first move to interaction picture associate with $H_0$. The control Hamiltonian becomes $H_c=\Omega(|1\rangle_A \langle d_A|\otimes U+{\rm H.c.})$. The ancilla dephasing errors $\{\sqrt{\kappa_m}|m\rangle_A\langle m|\otimes \mathbb{I}_B\}_{m=1}^{d_A}$ and the ancilla relaxation errors $\{\sqrt{\gamma_m}|m-1\rangle_A\langle m|\otimes \mathbb{I}_B\}_{m=3}^{d_A}$ remain unchanged, but the relaxation error $\sqrt{\gamma_1}|1\rangle_A\langle 2|\otimes \mathbb{I}_B$ becomes $\sqrt{\gamma_1}|1\rangle_A\langle 2|\otimes R_1^{\dagger}(t)R_2(t)$.

Obviously the control Hamiltonian and the ancilla dephasing errors are in the self-ajoint PI matrix subalgebra with the basis
\begin{eqnarray}\label{}
    \{|m\rangle_A\langle m|\otimes \mathbb{I}_B\}_{m=1}^{d_A}\cup\{|1\rangle_A\langle d_A|\otimes U, |d_A\rangle_A\langle 1|\otimes U^{\dagger}\},\nonumber \\
\end{eqnarray}
so if there are only ancilla dephasing errors, the PI gate condition is exactly satisfied.
If there are also ancilla relaxation errors $\{\sqrt{\gamma_m}|m-1\rangle_A\langle m|\otimes \mathbb{I}_B\}_{m=3}^{d_A}$, we can still form a larger self-adjoint PI matrix subalgebra with the basis
\begin{eqnarray}\label{}
    &\{|m\rangle_A\langle m|\otimes \mathbb{I}_B\}_{m=1}^{d_A}\cup\{|1\rangle_A\langle d_A|\otimes U, |d_A\rangle_A\langle 1|\otimes U^{\dagger}\} \nonumber \\
    &\cup \{|m-1\rangle_A\langle m|\otimes \mathbb{I}_B, |m\rangle_A\langle m-1|\otimes \mathbb{I}_B\}_{m=3}^{d_A},
\end{eqnarray}
but the addition of the relaxation error $\sqrt{\gamma_1}|1\rangle_A\langle 2|\otimes R_1^{\dagger}(t)R_2(t)$ destroys such a PI matrix subalgebra. For the diagram in Fig. \ref{example}(b), the PI gate condition is satisfied up to the $2$th-order from $|1\rangle_A$ to $|4\rangle_A$, the $3$th-order from $|1\rangle_A$ to $|3\rangle_A$, and the $4$th-order from $|1\rangle_A$ to $|2\rangle_A$ \cite{Ma2020}.


\begin{thebibliography}{}
\bibitem{footnote2} The reason we choose $H_0(t)$ to be diagonal in the ancilla eigenbasis is that it preserves the ancilla projection basis. When $H_0(t)$ contains off-diagonal terms in the ancilla eigenbasis (acting nontrivially on the central system), generallly it is not possible to guarantee a unitary operation on the central system when returning to the Schr\"{o}dinger's picture.
\end{thebibliography}
\end{document}